\documentclass[11pt,a4paper,twocolumn,nofootinbib]{revtex4}

\usepackage[cp1250]{inputenc}
\usepackage{graphicx} 
\usepackage{subfigure}
\usepackage{geometry}
\usepackage{amsmath}

\usepackage{color}  

\newgeometry{tmargin=2cm, bmargin=2cm, lmargin=2cm, rmargin=2cm}

\begin{document}

\title{Statistical mechanics of coevolving spin system}
\date{\today}
\author{Tomasz Raducha\textsuperscript{1,2}}
\email{tomasz.raducha@fuw.edu.pl}
\author{Mateusz Wilinski\textsuperscript{1,3}}
\email{mateusz.wilinski@fuw.edu.pl}
\author{Tomasz Gubiec\textsuperscript{1,4}}
\author{H. Eugene Stanley\textsuperscript{4}}
\affiliation{\textsuperscript{1}Institute of Experimental Physics,
  Faculty of Physics, University of Warsaw, Pasteura 5, 02-093 Warsaw,
  Poland} 
\affiliation{\textsuperscript{2}Instituto de Fisica Interdisciplinary
  Sistemas Complejos IFISC (CSIC-UIB), 07122 Palma de Mallorca, Spain} 
\affiliation{\textsuperscript{3}Scuola Normale Superiore, Piazza dei
  Cavalieri 7, 56126 Pisa, Italy} 
\affiliation{\textsuperscript{4}Center for Polymer Studies and
  Department of Physics, Boston University, Boston, Massachusetts 02215,
  USA} 

\begin{abstract}

We propose a statistical mechanics approach to a coevolving spin system with an adaptive network of interactions.  The dynamics of node states and network connections is driven by both spin configuration and network topology.  We consider a Hamiltonian that merges the classical Ising
model and the statistical theory of correlated random networks.  As a
result, we obtain rich phase diagrams with different phase transitions
both in the state of nodes and in the graph topology.  We argue that the
coupling between the spin dynamics and the structure of the network is
crucial in understanding the complex behavior of real-world systems and
omitting one of the approaches renders the description incomplete.

\end{abstract}

\maketitle


During the last few decades, there has been a rapid development in the interdisciplinary area of network science. This may be because of the availability of vast amounts of data, much of it from such complex systems as financial markets, social and biological structures, and transportation networks.  Studies of the network structure of such real-world systems as the World Wide Web \cite{albert1999internet}
indicate that their topology has numerous non-trivial properties that
the classical random graph model cannot explain \cite{erdos1959random}.
This has produced new network models able to recreate some of these
observed phenomena \cite{watts1998collective,barabasi1999emergence}.
Initially, most of these models focused on the graph evolution, often
the growth in the number of nodes and edges
\cite{dorogovtsev2002evolution}.  On the other hand, a different approach has been developed that considers a statistical ensemble of
graphs \cite{burda2001statistical} called ``exponential random graphs''
\cite{newmannetworks}.  This formalism, borrowed from statistical physics, has proved successful and has led to a phenomenological theory
of the topological phase transitions in evolving networks
\cite{burda2003uncorrelated,berg2002correlated,palla2004statistical}.

This newly discovered concept of networks with a complex structure moved rapidly through the spin models community.  Important critical properties were observed for both scale-free and small-world network versions of the canonical Ising model of ferromagnetism
\cite{herrero2002ising,aleksiejuk2002ferromagnetic,tadic2005magnetization}.
The use of complex networks became popular because they more closely
resemble real-world structures than regular lattices or Poissonian
graphs.  This has been particularly important when modeling social and
financial phenomena for which spin models are the simplest and the most
common \cite{sznajd2000opinion,cont2000herd}.  Although the topology of
many systems can be described using complex networks, the evolution of
the model is limited to changes in the spin configuration.  This implies
that connection dynamics evolve more slowly than node state dynamics.
Unfortunately, this assumption is not valid in most complex adaptive systems, which are describable using network tools. An Ising model with slowly evolving interactions was used as a model of a neural network
\cite{coolen1993coupled} and as a possible tool for simulating
magnetostriction in nanoscale magnetic structures
\cite{allahverdyan2006statistical}. A particularly interesting case
involved models in which the connections and state dynamics coevolve with each other, one evolution depending on the other and resulting in non-trivial feedback.  Most of these models focus on socio-economic systems and describe their dynamics \cite{mandra2009coevolution} rather than their statistical mechanics
\cite{biely2009socio,toruniewska2016unstable}.  Some of them produce intriguing topological properties, mainly when the dynamics is driven by the structural characteristics of the network
\cite{raducha2016coevolving}.


We here use the Hamiltonian formalism to describe Ising-like models with coevolution of spins and connections, and we want the connection dynamics to depend on both the spin configuration and network topology.
Following the approach taken in Ref.~\cite{berg2002correlated} we use
the degree as a topological variable and focus on nearest neighbor
interactions.  We consider undirected graphs with a fixed number $N$ of
vertices and a fixed number $M$ of edges. The partition function $Z$ for
our ensemble we define to be
\begin{equation} \label{eq:partition}
Z = \sum_{\{c_{ij}\},\{s_i\}} e^{-\beta H(\{c_{ij}\},\{s_i\})},
\end{equation}
where $\{\cdot\}$ is all possible configurations with respect to a fixed
number of links and nodes. Parameter $\beta$ is the strength of
fluctuations and is the inverse temperature.  A general form of the
Hamiltonian that lies within the scope of this paper is
\begin{equation}
\begin{split}
&H(\{c_{ij}\},\{s_i\}) =\\
&\sum_{i<j} c_{ij} f(k_i,k_j,s_i,s_j) + \sum_{i} g(k_i,s_i),
\end{split}
\end{equation}
where $c_{ij}$ is the adjacency matrix, $k_i = \sum_j c_{ij}$ and $s_i$
are respectively the degree and the spin of node $i$, and $f(\cdot)$ and
$g(\cdot)$ are functions to be determined.  More specifically, we assume
that the functions are such that
\begin{equation}
\begin{split}
&H(\{c_{ij}\},\{s_i\}) =\\
&- \sum_{i<j} c_{ij} \left( \frac{k_i k_j}{\langle k \rangle}
  \right)^{\phi} s_i s_j - \sum_{i} k_i^{\gamma}  - h \sum_{i} s_i, 
\end{split}
\end{equation}
where $\phi$ and $\gamma$ are model parameters, $\langle k \rangle =
2M/N$, and $h$ is the external field acting on spins set to zero. This
simple concrete form shows (i) that parameters $\phi$ and $\gamma$ allow
us to continuously switch from complicated topological interactions to
the classical Ising model, and (ii) that the multiplication of degrees
is the simplest interaction expression.  In an Ising framework, we treat
it as a weight $J_{ij}$ assigned to an edge $(i,j)$.  In addition, $J_{ij} = \left(\frac{k_i k_j}{\langle k \rangle}\right)^{\phi}$ is in
accordance with real-world weighted network characteristics
\cite{barrat2004architecture}.  The second sum term is an external field
that interacts with each local node degree and drives the preference for
high or low degree nodes. In addition to the classical ferromagnetic
interpretation, if we use a socio-economic model to examine the proposed
Hamiltonian we find an accurate interpretation of its terms. Using an
opinion model we determine the influence of a given agent by examining
its connectivity.  The external field term forces each agent to reach as
many people as possible. In contrast, the interaction term allows the
energy of the system to be strongly affected by the connections among
influential high degree nodes.  This works in two ways. High-degree
agents with opposite spins are energetically unstable, and agents with the same spins lower the energy level.


\begin{figure}
\includegraphics[width=1.05\linewidth]{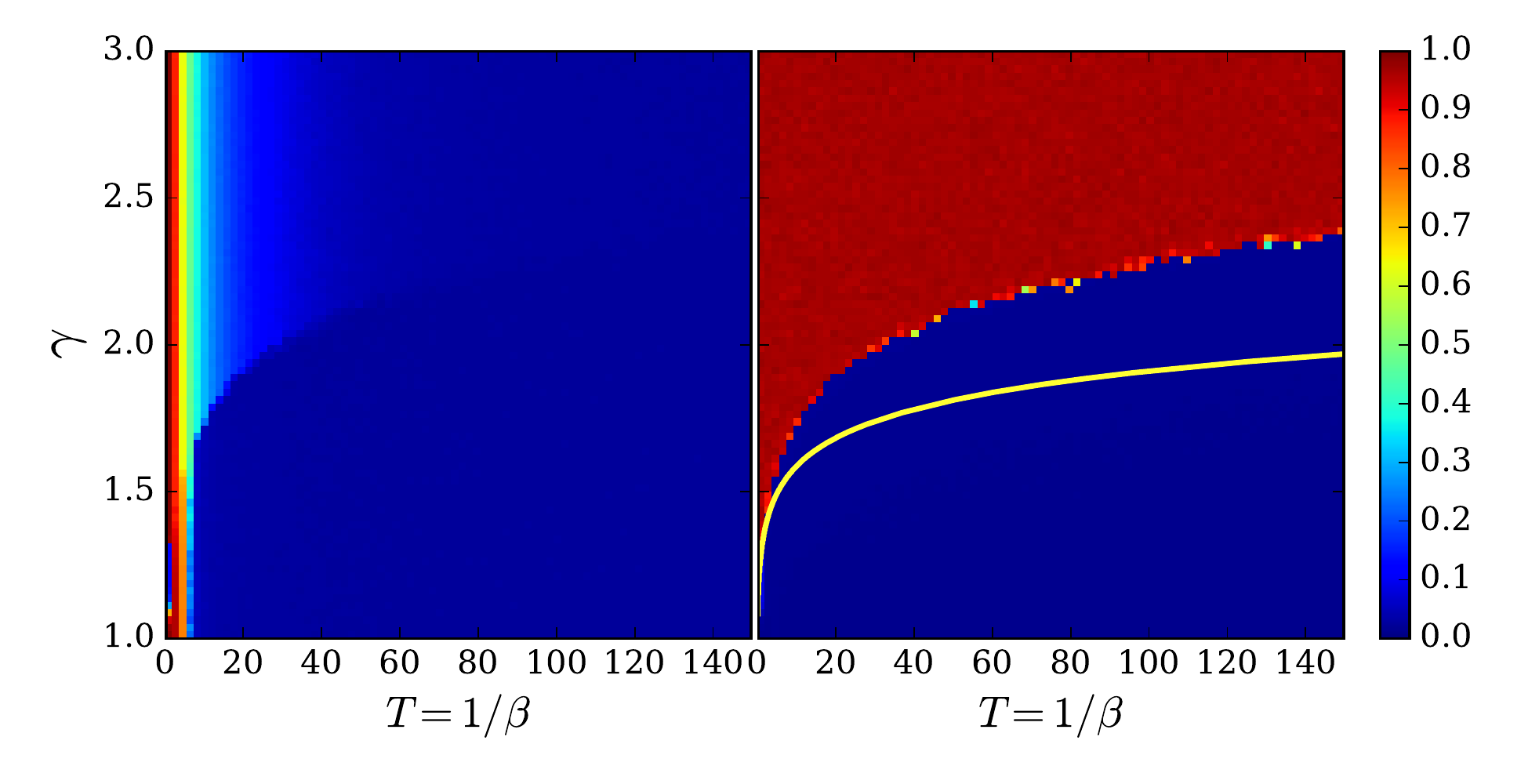}
\caption{(Color online) Absolute magnetization $|m|$ (left) and the largest degree
  $k_{max}$ (right), as a function of the temperature $T = 1/\beta$ and
  $\gamma$, for $\phi = 0$.  Solid line represents analytical
  approximation of the transition according to the equation
  (\ref{eq:Z_star}).  Results averaged over $5 \cdot 10^5$ time steps
  for a network with $N=1000$ nodes and $M=3000$ edges.}
\label{fig:2D_plot_gamma}
\end{figure}

\begin{figure}
  \includegraphics[width=1\linewidth]{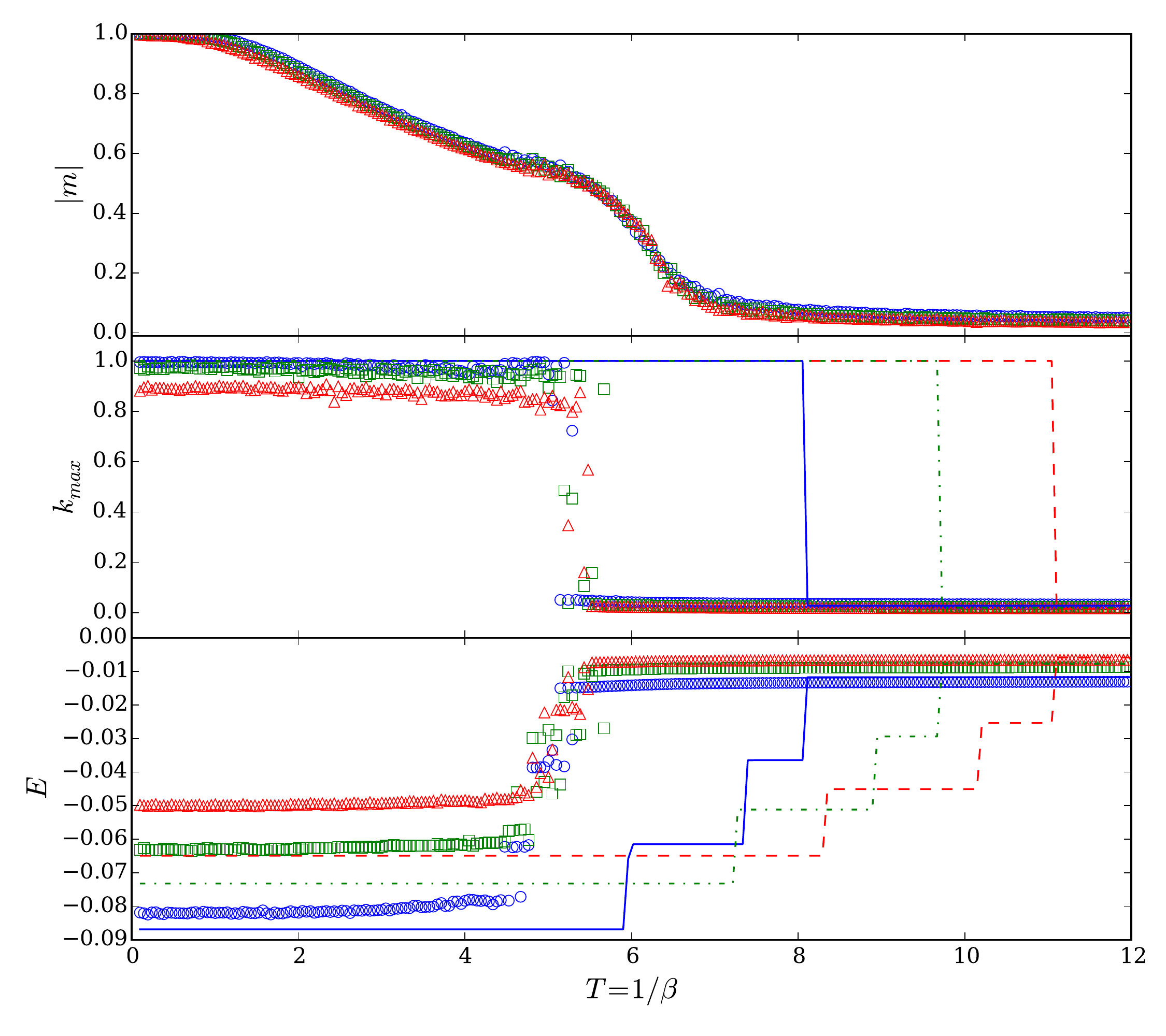}
\caption{(Color online) Absolute magnetization $|m|$, the largest degree $k_{max}$,
  and energy $E$ as a function of the temperature $T=1/\beta$, for
  $\gamma = 1.6$ and $\phi = 0$.  Lines represent analytical
  approximations according to the equation (\ref{eq:Z_star}) and symbols
  correspond to numerical simulations averaged over $10^6$ time steps
  for a network with $N=500$ (blue circles and solid line), $N=750$
  (green squares and dotted line), $N=1000$ (red triangles and dashed
  line), with $c=M/N=3$.  All quantities normalized to the range
  $[0,1]$, except the energy, which is given in arbitrary units.}
\label{fig:1D_plot_gamma}
\end{figure}

The topological portion of the Hamiltonian changes the behavior of the Ising model. We find a variety of different effects.  Some are structural, others are associated with spin configurations, and still others are a result of both. Figure~\ref{fig:2D_plot_gamma} shows the
simulation results for the phase diagram of $\gamma$ and $T$ with a
fixed $\phi = 0$. Here, no structural portion of the  Hamiltonian describes interactions, i.e., the network structure is only locally important. Note that there are two separate topological phases and also
a continuous phase transition in magnetization, with a small impact by
parameter $\gamma$.

\begin{figure}[h]
\subfigure[Model: $\phi = 0$, $\gamma = 2$]{
 \includegraphics[width=0.46\linewidth]{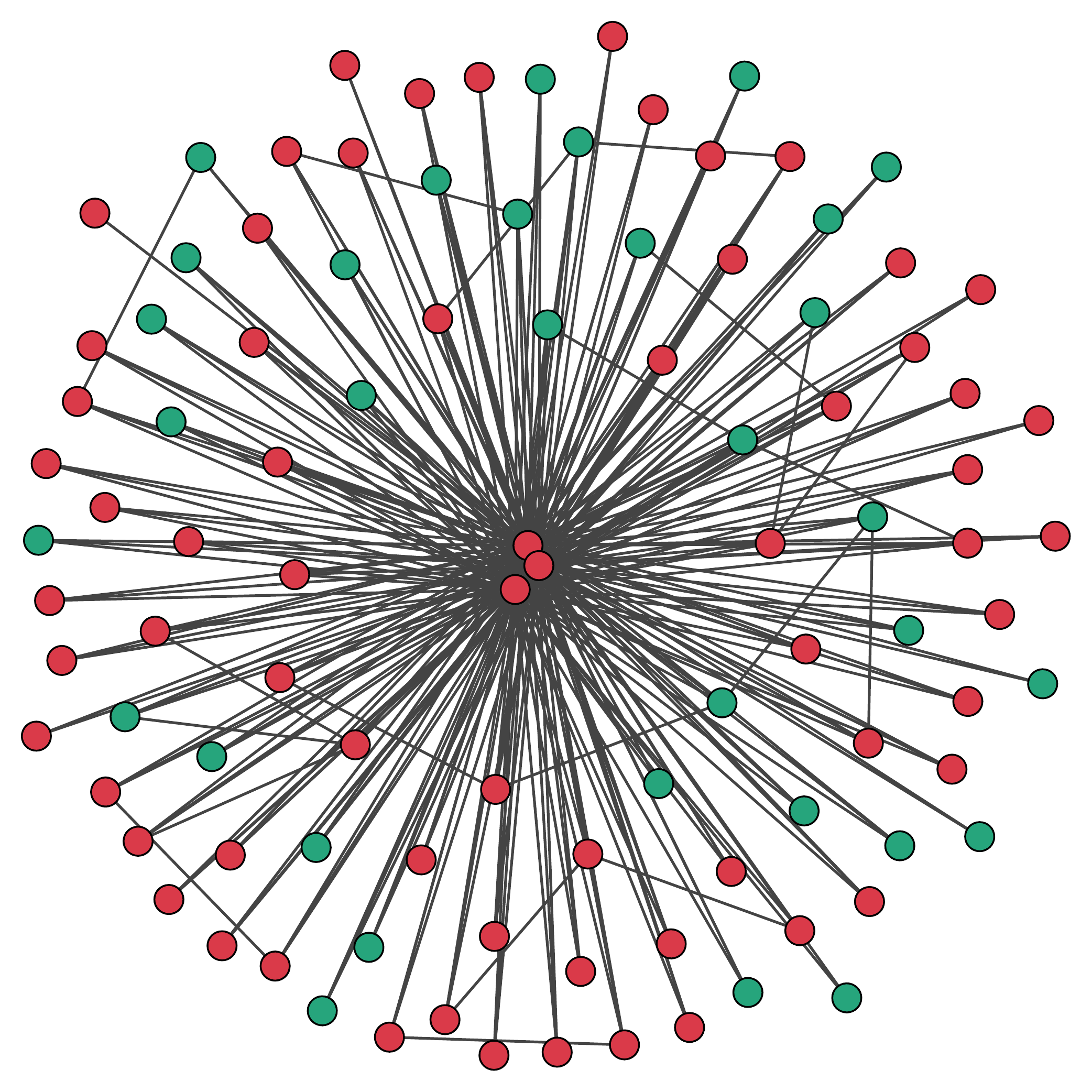}
}
\subfigure[Model: $\phi = 1$, $\gamma = 1$]{
 \includegraphics[width=0.46\linewidth]{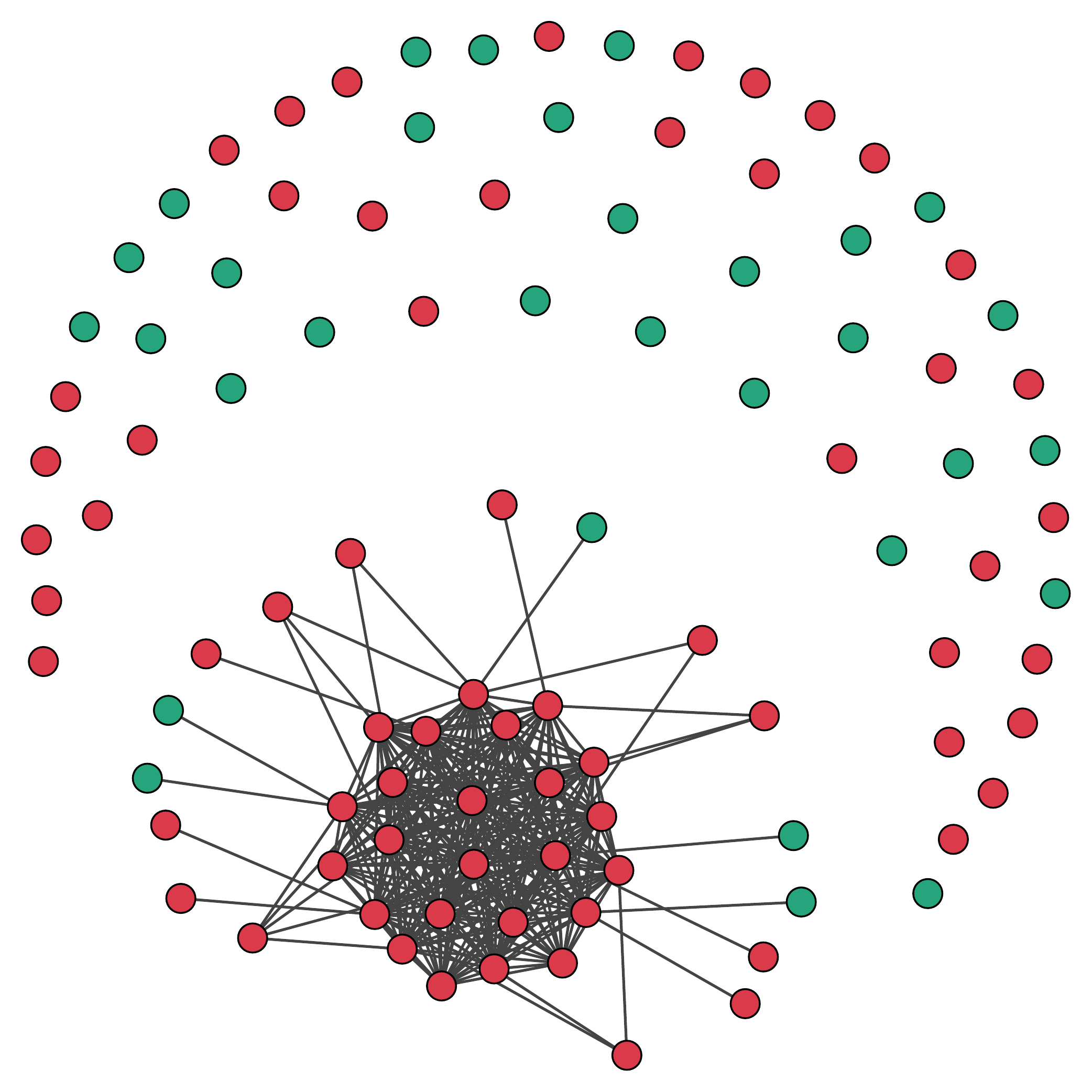}
}
\caption{(Color online) Exemplary networks obtained in the simulation.  Green nodes
  indicate $+1$ spin, red nodes indicate $-1$ spin.}
\label{fig:graph}
\end{figure}

Figure~\ref{fig:1D_plot_gamma} shows the case $\phi = 0$ and $\gamma =
1.6$. Note that the topological transition of the highest degree is
discontinuous, but also that the magnetization behaves in a way similar
to a standard Ising model on a coevolving network
\cite{biely2009socio}. These effects belong to different transition
classes and occur at different temperatures, and we see a striking behavior in energy $E$, i.e., the value of the Hamiltonian. It exhibits
multiple jumps, one of which occurs at the same temperature as the
highest-degree topological phase transition. In addition, all jumps are approximately equal.  This energy behavior suggests a multi-star configuration in which the maximum number of stars is restricted by a fraction $\frac{M}{N}$. Figure~\ref{fig:graph}(a) shows that when $M = 3N$ three vertices are connected to approximately every network node.
The number of stars decreases with the temperature. Eventually, the
system becomes more homogeneous, and we see a sharp transition in the largest degree. Here $k_{\rm max} \ll N$, and the degree distribution is approximately Poissonian.

\begin{figure}
\includegraphics[width=1.05\linewidth]{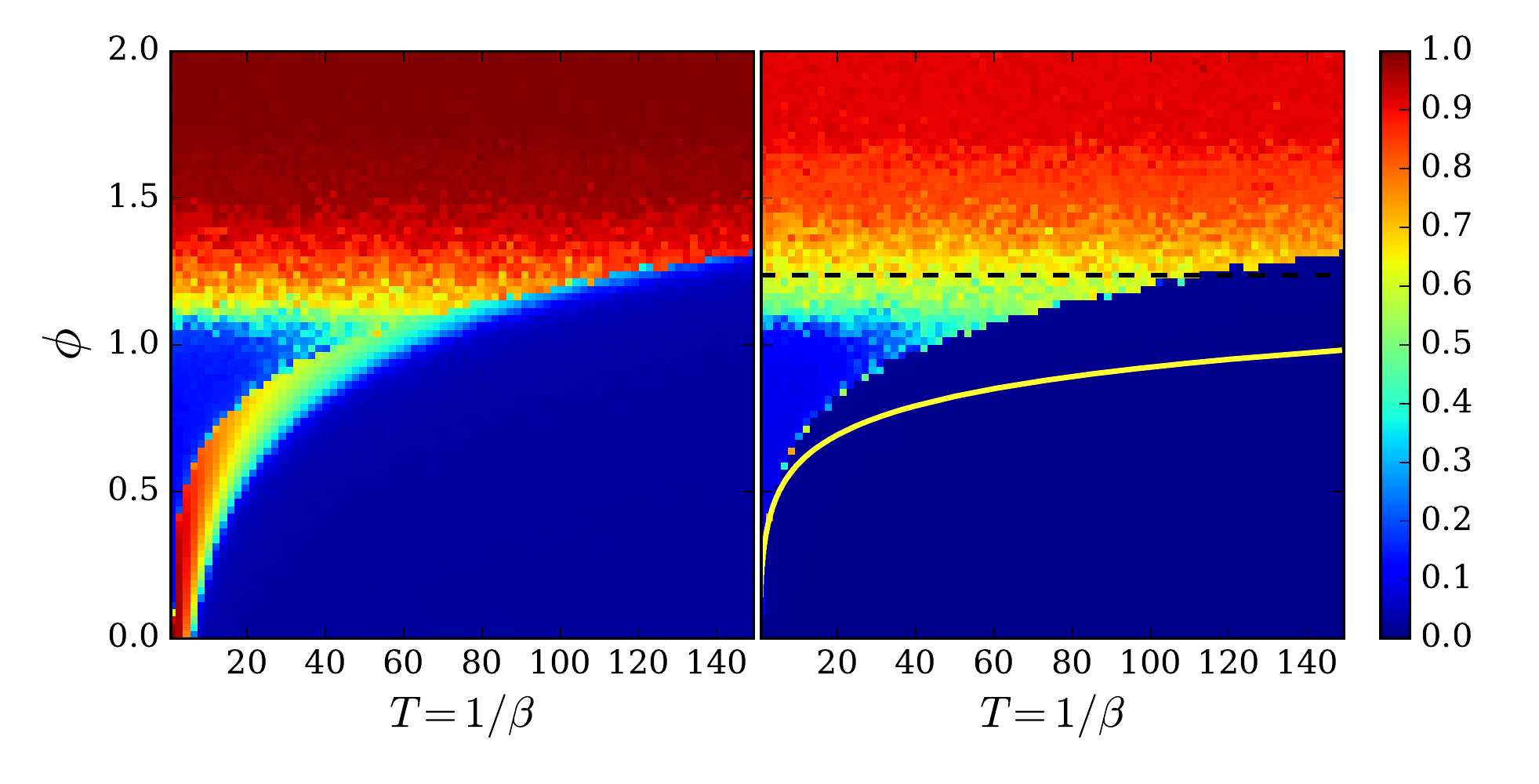}
\caption{(Color online) Absolute magnetization $|m|$ (left) and the largest degree
  $k_{max}$ (right), as a function of the temperature $T = 1/\beta$ and
  $\phi$, for $\gamma = 1$.  Solid and dashed lines represent analytical
  approximations of the transitions according to equations
  (\ref{eq:Z_complete_g}) and (\ref{eq:phi_c}) respectively.  Results
  averaged over $5 \cdot 10^5$ time steps for a network with $N=1000$
  nodes and $M=3000$ edges.}
\label{fig:2D_plot_phi}
\end{figure}

\begin{figure}
  \includegraphics[width=1\linewidth]{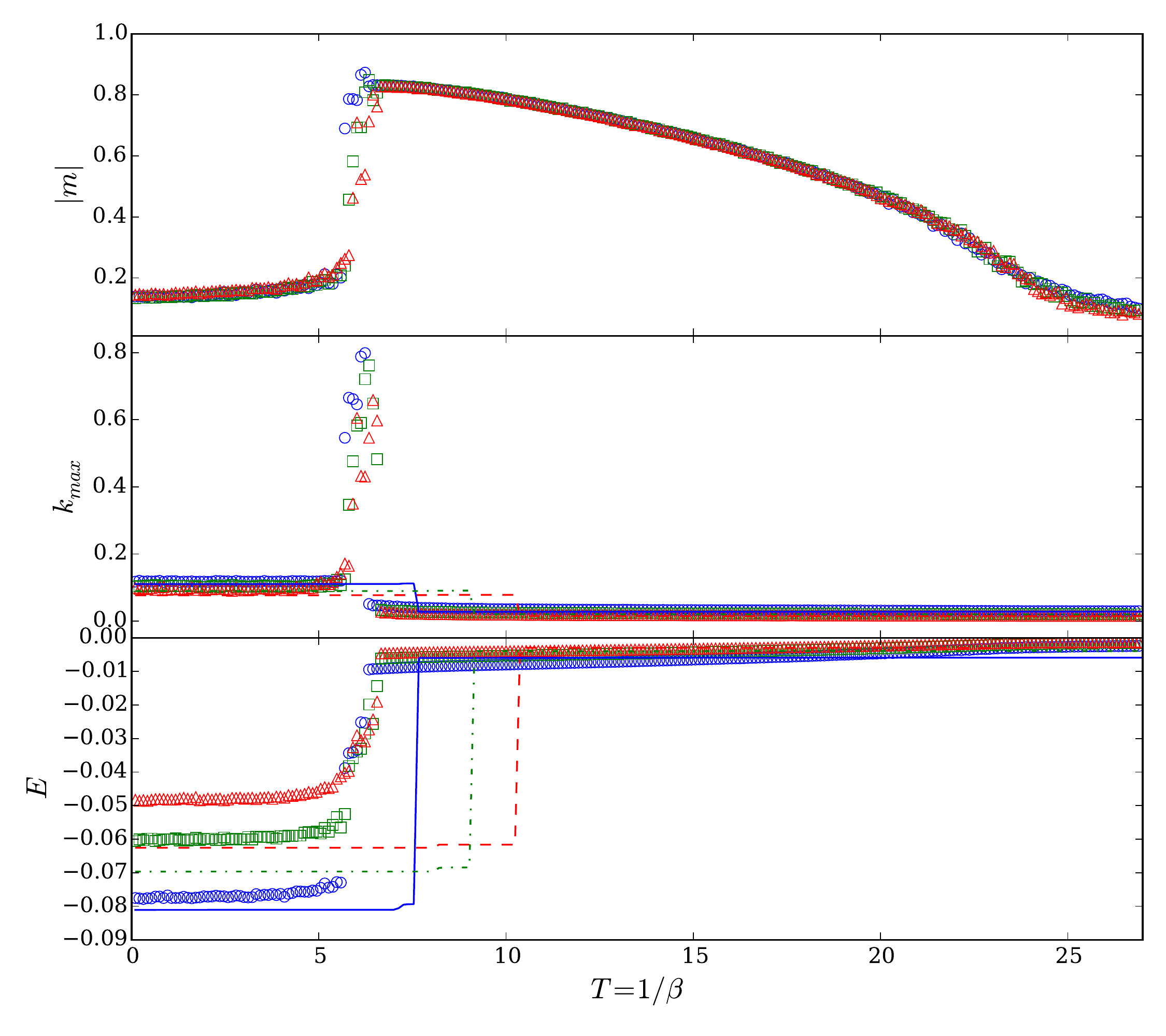}
\caption{(Color online) Absolute magnetization $|m|$, the largest degree $k_{max}$,
  and energy $E$ as a function of the temperature $T=1/\beta$, for $\phi
  = 0.6$ and $\gamma = 1$.  Lines represent analytical approximations
  according to the equation (\ref{eq:Z_complete_g}) and symbols
  correspond to numerical simulations averaged over $10^6$ time steps
  for a network with $N=500$ (blue circles and solid line), $N=750$
  (green squares and dotted line), $N=1000$ (red triangles and dashed
  line), with $c=M/N=3$.  All quantities normalized to the range
  $[0,1]$, except the energy, which is given in arbitrary units.}
\label{fig:1D_plot_phi}
\end{figure}

When we remove the external field associated with the degree of each
node and turn on the combination of structural terms in the interaction portion of the Hamiltonian, we find a different
behavior. Figure~\ref{fig:2D_plot_phi} shows this in the phase diagram with respect to $\phi$ and $T$ when there is a neutral value of $\gamma = 1$. Figure~\ref{fig:1D_plot_phi} shows the same when we fix $\phi = 0.6$. When we examine the largest degree and the magnetization we see four phases. Figure~\ref{fig:1D_plot_phi} shows that in the one-dimensional phase diagram the topological transition is
characterized by a sharp jump in the maximum degree. Thus there is an
abrupt change in the magnetization. Unlike the case with a varying
$\gamma$, there is also a transition that is triggered by a change in
parameter $\phi$, and that is unaffected by temperature. The critical
value of $\phi$ in this transition is notated $\phi_c$.

Examining the structural properties of the different phases mentioned above, we find that when $\gamma = 1$ and $\phi < \phi_c$ in a low-temperature regime there are many disconnected nodes and one big component with high-degree clustering [see Fig.~\ref{fig:graph}(b)].  At the critical temperature, the network recombines into one component, and the highest-degree $k_{\rm max}$ reaches its maximum value at the transition point. Increasing the temperature decreases the highest degree and the degree distribution to become Poissonian.  When we increase $\phi$ above $\phi_c$, the system transitions into a multi-star configuration, a phase similar to the one previously observed for high
$\gamma$ values.

We next analytically describe the system to produce an approximation of our numerical results. Because the structural heterogeneity disallows a simple mean-field approach, we use a semi-mean-field method and focus on the non-homogeneous elements of the system that most strongly impact the Hamiltonian.

Figures \ref{fig:2D_plot_gamma} and \ref{fig:1D_plot_gamma} show the
results when $\phi = 0$ and the $\gamma$ value varies. We here assume
that the most important part of the Hamiltonian is the contribution from
the largest hubs, i.e., the stars connected to all other nodes. We
assume non-hubs to have a degree equal to the average degree.  Thus we
approximate the Hamiltonian
\begin{equation} \label{eq:H_semi-mean}
H \approx - n_h k_{m}^{\gamma} - (N-n_h)k^{\gamma}_a,
\end{equation}
where $n_h \in \{0, 1, 2, ... , \lfloor \frac{M}{N} \rfloor \equiv
\lfloor c \rfloor \}$ is the number of stars, $k_{m} = N-1$ is the
degree of each star, $k_a$ is the average degree of the remaining nodes,
and $k_a \equiv k_a(n_h) = n_h+\frac{2 (M - L(n_h))}{N-n_h}$, where
$L(n_h) = k_m + k_m-1 + ... + k_m - (n_h-1) = n_h \frac{2N-1-n_h}{2}$ is
the number of links required to create $n_h$ stars.  Taking into account
all possible configurations, the partition function becomes
\begin{equation}\label{eq:Z_star}
Z = 2^N  \sum_{n_h=0}^{\lfloor c \rfloor} \binom{N}{n_h}
R(n_h) e^{\beta[ n_h k_m^\gamma + (N-n_h)k^{\gamma}_a]},
\end{equation}
where $2^N$ is all spin configurations, $\binom{N}{n_h}$ is the number
of star combinations, and $R(n_h)$ is the number of possible link
configurations with $n_h$ stars in the network.  We approximate this as
$R(n_h) \approx \binom{(N-n_h)(N-n_h-1)/2}{M-L(n_h)}$. Although this is
a slight over-counting when $n_h < c$, the number of incorrect
configurations is negligible when compared to the number of all other
configurations.

Figures \ref{fig:2D_plot_gamma} and \ref{fig:1D_plot_gamma} show that the partition function allows us to analytically determine the energy and the highest degree.  Although the estimated critical temperature diverges from the observed temperature, using this simple approach allows us to recreate the step-like behavior of the energy.

When $\gamma = 1$ and the $\phi < \phi_c$ value varies, there is a shattering transition with a decreasing temperature. Some nodes disconnect from other nodes and become inactive. In contrast, when the temperature is high, and $c=3$, the graph becomes random and highly connected. Thus we describe the state of the system in terms of the number of active nodes, and we assume that their degree can be approximated using the mean field approach. We denote the number of these nodes $n_s$ and write the Hamiltonian
\begin{equation}
H \approx - M \frac{ \langle k \rangle_s^{2\phi}}{ \langle k
  \rangle^{\phi}}, 
\end{equation}
where $\langle k \rangle_s = 2M/n_s$ and $\lceil n_{\rm min} \rceil \le
n_s \le N$ with $n_{\rm min} = (1+\sqrt{1+8M})/2$. We approximate the
number of configurations for a particular $n_s$ with $\binom{N}{n_s}
\binom{\frac{n_s(n_s-1)}{2}}{M}$ and derive the partition function
\begin{equation}\label{eq:Z_complete_g}
\begin{split}
Z &= \sum_{n_s = \lceil n_{\min} \rceil}^N 2^{N-n_s+1} \binom{N}{n_s} \times \\
&\times \binom{\frac{n_s(n_s-1)}{2}}{M} e^{\beta \, 2^{\phi} \, N^{\phi}
  \, M^{\phi + 1} \, n_s^{-2\phi}}.
\end{split}
\end{equation}
We assume that the spin direction of all active nodes is the same and
that there are $2^{N-n_s+1}$ possible spin
configurations. Figures~\ref{fig:2D_plot_phi} and \ref{fig:1D_plot_phi}
show the results when we analytically obtain the energy and the highest
degree level. As in the previous case, we can use our estimation to
approximate the system behavior, but not the critical temperature.

We approximate $\phi_c$ to fully describe the phase diagrams. The
critical value of $\phi$ separates the homogeneous active node phase
from the multi-star configuration phase. We assume that the energy of
both phases is the same when $\phi = \phi_c$ and define the critical
value
\begin{equation}\label{eq:phi_c}
\begin{split}
&\phi_c \ln \frac{(n_{\min}-1)^2}{N-1} + \ln \frac{M}{c} =\\
&\ln \left( \frac{c-1}{2} (N-1)^{\phi_c} + c^{\phi_c} (N-c) \right),
\end{split}
\end{equation}
where both $c$ and $n_{\rm min}$ retain the previous definitions.  For
the complete calculation of all cases see the Supplemental Material \cite{suppl}.


To statistically describe a coevolving spin system we have used a
Hamiltonian that merges exponential random graphs and Ising-like models.
A Hamiltonian that simultaneously depends on topological properties and node states has not been previously analyzed, and we have found complex behavior and have generated rich phase diagrams.  The most striking aspect of our results is the existence, at specific temperatures, of topological phase transitions in which there are no node state transitions. There are also transitions that influence order parameters, but this suggests that we must take into account both the topology and the state of the nodes to fully describe the system, and if we do not we miss essential aspects of systemic behavior.

Although the results presented here concern networks in which $c \equiv \frac{M}{N} = 3$, we did extend our simulations to other cases and found the same qualitative results.  These extended results and a detailed analysis of the asymptotic and topological properties of the transitions will be supplied in a future publication \cite{inpreparation}.


\textbf{Acknowledgments:} Mateusz Wilinski would like to acknowledge the
support from the National Science Centre under Grant
2015/19/N/ST2/02701.  Tomasz Raducha and Mateusz Wilinski would like to acknowledge the support from the Young Researchers of the Complex Systems Society under Bridge Grant 2018.


\begin{thebibliography}{24}
\expandafter\ifx\csname natexlab\endcsname\relax\def\natexlab#1{#1}\fi
\expandafter\ifx\csname bibnamefont\endcsname\relax
  \def\bibnamefont#1{#1}\fi
\expandafter\ifx\csname bibfnamefont\endcsname\relax
  \def\bibfnamefont#1{#1}\fi
\expandafter\ifx\csname citenamefont\endcsname\relax
  \def\citenamefont#1{#1}\fi
\expandafter\ifx\csname url\endcsname\relax
  \def\url#1{\texttt{#1}}\fi
\expandafter\ifx\csname urlprefix\endcsname\relax\def\urlprefix{URL }\fi
\providecommand{\bibinfo}[2]{#2}
\providecommand{\eprint}[2][]{\url{#2}}

\bibitem[{\citenamefont{Albert et~al.}(1999)\citenamefont{Albert, Jeong, and
  Barab{\'a}si}}]{albert1999internet}
\bibinfo{author}{\bibfnamefont{R.}~\bibnamefont{Albert}},
  \bibinfo{author}{\bibfnamefont{H.}~\bibnamefont{Jeong}}, \bibnamefont{and}
  \bibinfo{author}{\bibfnamefont{A.-L.} \bibnamefont{Barab{\'a}si}},
  \bibinfo{journal}{nature} \textbf{\bibinfo{volume}{401}},
  \bibinfo{pages}{130} (\bibinfo{year}{1999}).

\bibitem[{\citenamefont{Erd{\"o}s and R{\'e}nyi}(1959)}]{erdos1959random}
\bibinfo{author}{\bibfnamefont{P.}~\bibnamefont{Erd{\"o}s}} \bibnamefont{and}
  \bibinfo{author}{\bibfnamefont{A.}~\bibnamefont{R{\'e}nyi}},
  \bibinfo{journal}{Publicationes Mathematicae (Debrecen)}
  \textbf{\bibinfo{volume}{6}}, \bibinfo{pages}{290} (\bibinfo{year}{1959}).

\bibitem[{\citenamefont{Watts and Strogatz}(1998)}]{watts1998collective}
\bibinfo{author}{\bibfnamefont{D.~J.} \bibnamefont{Watts}} \bibnamefont{and}
  \bibinfo{author}{\bibfnamefont{S.~H.} \bibnamefont{Strogatz}},
  \bibinfo{journal}{nature} \textbf{\bibinfo{volume}{393}},
  \bibinfo{pages}{440} (\bibinfo{year}{1998}).

\bibitem[{\citenamefont{Barab{\'a}si and Albert}(1999)}]{barabasi1999emergence}
\bibinfo{author}{\bibfnamefont{A.-L.} \bibnamefont{Barab{\'a}si}}
  \bibnamefont{and} \bibinfo{author}{\bibfnamefont{R.}~\bibnamefont{Albert}},
  \bibinfo{journal}{science} \textbf{\bibinfo{volume}{286}},
  \bibinfo{pages}{509} (\bibinfo{year}{1999}).

\bibitem[{\citenamefont{Dorogovtsev and
  Mendes}(2002)}]{dorogovtsev2002evolution}
\bibinfo{author}{\bibfnamefont{S.~N.} \bibnamefont{Dorogovtsev}}
  \bibnamefont{and} \bibinfo{author}{\bibfnamefont{J.~F.}
  \bibnamefont{Mendes}}, \bibinfo{journal}{Advances in physics}
  \textbf{\bibinfo{volume}{51}}, \bibinfo{pages}{1079} (\bibinfo{year}{2002}).

\bibitem[{\citenamefont{Burda et~al.}(2001)\citenamefont{Burda, Correia, and
  Krzywicki}}]{burda2001statistical}
\bibinfo{author}{\bibfnamefont{Z.}~\bibnamefont{Burda}},
  \bibinfo{author}{\bibfnamefont{J.~D.} \bibnamefont{Correia}},
  \bibnamefont{and}
  \bibinfo{author}{\bibfnamefont{A.}~\bibnamefont{Krzywicki}},
  \bibinfo{journal}{Physical Review E} \textbf{\bibinfo{volume}{64}},
  \bibinfo{pages}{046118} (\bibinfo{year}{2001}).

\bibitem[{\citenamefont{Newman}(2016)}]{newmannetworks}
\bibinfo{author}{\bibfnamefont{M.}~\bibnamefont{Newman}}, pp.
  \bibinfo{pages}{565--588} (\bibinfo{year}{2016}).

\bibitem[{\citenamefont{Burda and Krzywicki}(2003)}]{burda2003uncorrelated}
\bibinfo{author}{\bibfnamefont{Z.}~\bibnamefont{Burda}} \bibnamefont{and}
  \bibinfo{author}{\bibfnamefont{A.}~\bibnamefont{Krzywicki}},
  \bibinfo{journal}{Physical Review E} \textbf{\bibinfo{volume}{67}},
  \bibinfo{pages}{046118} (\bibinfo{year}{2003}).

\bibitem[{\citenamefont{Berg and L{\"a}ssig}(2002)}]{berg2002correlated}
\bibinfo{author}{\bibfnamefont{J.}~\bibnamefont{Berg}} \bibnamefont{and}
  \bibinfo{author}{\bibfnamefont{M.}~\bibnamefont{L{\"a}ssig}},
  \bibinfo{journal}{Physical review letters} \textbf{\bibinfo{volume}{89}},
  \bibinfo{pages}{228701} (\bibinfo{year}{2002}).

\bibitem[{\citenamefont{Palla et~al.}(2004)\citenamefont{Palla, Der{\'e}nyi,
  Farkas, and Vicsek}}]{palla2004statistical}
\bibinfo{author}{\bibfnamefont{G.}~\bibnamefont{Palla}},
  \bibinfo{author}{\bibfnamefont{I.}~\bibnamefont{Der{\'e}nyi}},
  \bibinfo{author}{\bibfnamefont{I.}~\bibnamefont{Farkas}}, \bibnamefont{and}
  \bibinfo{author}{\bibfnamefont{T.}~\bibnamefont{Vicsek}},
  \bibinfo{journal}{Physical Review E} \textbf{\bibinfo{volume}{69}},
  \bibinfo{pages}{046117} (\bibinfo{year}{2004}).

\bibitem[{\citenamefont{Herrero}(2002)}]{herrero2002ising}
\bibinfo{author}{\bibfnamefont{C.~P.} \bibnamefont{Herrero}},
  \bibinfo{journal}{Physical Review E} \textbf{\bibinfo{volume}{65}},
  \bibinfo{pages}{066110} (\bibinfo{year}{2002}).

\bibitem[{\citenamefont{Aleksiejuk et~al.}(2002)\citenamefont{Aleksiejuk,
  Ho{\l}yst, and Stauffer}}]{aleksiejuk2002ferromagnetic}
\bibinfo{author}{\bibfnamefont{A.}~\bibnamefont{Aleksiejuk}},
  \bibinfo{author}{\bibfnamefont{J.~A.} \bibnamefont{Ho{\l}yst}},
  \bibnamefont{and} \bibinfo{author}{\bibfnamefont{D.}~\bibnamefont{Stauffer}},
  \bibinfo{journal}{Physica A: Statistical Mechanics and its Applications}
  \textbf{\bibinfo{volume}{310}}, \bibinfo{pages}{260} (\bibinfo{year}{2002}).

\bibitem[{\citenamefont{Tadi{\'c} et~al.}(2005)\citenamefont{Tadi{\'c}, Malarz,
  and Ku{\l}akowski}}]{tadic2005magnetization}
\bibinfo{author}{\bibfnamefont{B.}~\bibnamefont{Tadi{\'c}}},
  \bibinfo{author}{\bibfnamefont{K.}~\bibnamefont{Malarz}}, \bibnamefont{and}
  \bibinfo{author}{\bibfnamefont{K.}~\bibnamefont{Ku{\l}akowski}},
  \bibinfo{journal}{Physical review letters} \textbf{\bibinfo{volume}{94}},
  \bibinfo{pages}{137204} (\bibinfo{year}{2005}).

\bibitem[{\citenamefont{Sznajd-Weron and Sznajd}(2000)}]{sznajd2000opinion}
\bibinfo{author}{\bibfnamefont{K.}~\bibnamefont{Sznajd-Weron}}
  \bibnamefont{and} \bibinfo{author}{\bibfnamefont{J.}~\bibnamefont{Sznajd}},
  \bibinfo{journal}{International Journal of Modern Physics C}
  \textbf{\bibinfo{volume}{11}}, \bibinfo{pages}{1157} (\bibinfo{year}{2000}).

\bibitem[{\citenamefont{Cont and Bouchaud}(2000)}]{cont2000herd}
\bibinfo{author}{\bibfnamefont{R.}~\bibnamefont{Cont}} \bibnamefont{and}
  \bibinfo{author}{\bibfnamefont{J.-P.} \bibnamefont{Bouchaud}},
  \bibinfo{journal}{Macroeconomic dynamics} \textbf{\bibinfo{volume}{4}},
  \bibinfo{pages}{170} (\bibinfo{year}{2000}).

\bibitem[{\citenamefont{Coolen et~al.}(1993)\citenamefont{Coolen, Penney, and
  Sherrington}}]{coolen1993coupled}
\bibinfo{author}{\bibfnamefont{A.}~\bibnamefont{Coolen}},
  \bibinfo{author}{\bibfnamefont{R.}~\bibnamefont{Penney}}, \bibnamefont{and}
  \bibinfo{author}{\bibfnamefont{D.}~\bibnamefont{Sherrington}},
  \bibinfo{journal}{Physical Review B} \textbf{\bibinfo{volume}{48}},
  \bibinfo{pages}{16116} (\bibinfo{year}{1993}).

\bibitem[{\citenamefont{Allahverdyan and
  Petrosyan}(2006)}]{allahverdyan2006statistical}
\bibinfo{author}{\bibfnamefont{A.}~\bibnamefont{Allahverdyan}}
  \bibnamefont{and}
  \bibinfo{author}{\bibfnamefont{K.}~\bibnamefont{Petrosyan}},
  \bibinfo{journal}{EPL (Europhysics Letters)} \textbf{\bibinfo{volume}{75}},
  \bibinfo{pages}{908} (\bibinfo{year}{2006}).

\bibitem[{\citenamefont{Mandr{\`a} et~al.}(2009)\citenamefont{Mandr{\`a},
  Fortunato, and Castellano}}]{mandra2009coevolution}
\bibinfo{author}{\bibfnamefont{S.}~\bibnamefont{Mandr{\`a}}},
  \bibinfo{author}{\bibfnamefont{S.}~\bibnamefont{Fortunato}},
  \bibnamefont{and}
  \bibinfo{author}{\bibfnamefont{C.}~\bibnamefont{Castellano}},
  \bibinfo{journal}{Physical Review E} \textbf{\bibinfo{volume}{80}},
  \bibinfo{pages}{056105} (\bibinfo{year}{2009}).

\bibitem[{\citenamefont{Biely et~al.}(2009)\citenamefont{Biely, Hanel, and
  Thurner}}]{biely2009socio}
\bibinfo{author}{\bibfnamefont{C.}~\bibnamefont{Biely}},
  \bibinfo{author}{\bibfnamefont{R.}~\bibnamefont{Hanel}}, \bibnamefont{and}
  \bibinfo{author}{\bibfnamefont{S.}~\bibnamefont{Thurner}},
  \bibinfo{journal}{The European Physical Journal B-Condensed Matter and
  Complex Systems} \textbf{\bibinfo{volume}{67}}, \bibinfo{pages}{285}
  (\bibinfo{year}{2009}).

\bibitem[{\citenamefont{Toruniewska et~al.}(2016)\citenamefont{Toruniewska,
  Suchecki, and Ho{\l}yst}}]{toruniewska2016unstable}
\bibinfo{author}{\bibfnamefont{J.}~\bibnamefont{Toruniewska}},
  \bibinfo{author}{\bibfnamefont{K.}~\bibnamefont{Suchecki}}, \bibnamefont{and}
  \bibinfo{author}{\bibfnamefont{J.~A.} \bibnamefont{Ho{\l}yst}},
  \bibinfo{journal}{Physica A: Statistical Mechanics and its Applications}
  \textbf{\bibinfo{volume}{460}}, \bibinfo{pages}{1} (\bibinfo{year}{2016}).

\bibitem[{\citenamefont{Raducha and Gubiec}(2017)}]{raducha2016coevolving}
\bibinfo{author}{\bibfnamefont{T.}~\bibnamefont{Raducha}} \bibnamefont{and}
  \bibinfo{author}{\bibfnamefont{T.}~\bibnamefont{Gubiec}},
  \bibinfo{journal}{Physica A: Statistical Mechanics and its Applications}
  (\bibinfo{year}{2017}).

\bibitem[{\citenamefont{Barrat et~al.}(2004)\citenamefont{Barrat, Barthelemy,
  Pastor-Satorras, and Vespignani}}]{barrat2004architecture}
\bibinfo{author}{\bibfnamefont{A.}~\bibnamefont{Barrat}},
  \bibinfo{author}{\bibfnamefont{M.}~\bibnamefont{Barthelemy}},
  \bibinfo{author}{\bibfnamefont{R.}~\bibnamefont{Pastor-Satorras}},
  \bibnamefont{and}
  \bibinfo{author}{\bibfnamefont{A.}~\bibnamefont{Vespignani}},
  \bibinfo{journal}{Proceedings of the National Academy of Sciences of the
  United States of America} \textbf{\bibinfo{volume}{101}},
  \bibinfo{pages}{3747} (\bibinfo{year}{2004}).

\bibitem[{sup()}]{suppl}
\bibinfo{note}{See Supplemental Material at [URL to be inserted by publisher]
  for the formulas derivation and simulation details}.

\bibitem[{\citenamefont{Raducha and Wilinski}(In preparation)}]{inpreparation}
\bibinfo{author}{\bibfnamefont{T.}~\bibnamefont{Raducha}} \bibnamefont{and}
  \bibinfo{author}{\bibfnamefont{M.}~\bibnamefont{Wilinski}} (\bibinfo{year}{In
  preparation}).

\end{thebibliography}
\end{document}


\title{Supplemental Material for\\Statistical mechanics of coevolving
  spin system} 
\date{\today}
\author{Tomasz Raducha\textsuperscript{1,2}}
\email{tomasz.raducha@fuw.edu.pl}
\author{Mateusz Wilinski\textsuperscript{1,3}}
\email{mateusz.wilinski@fuw.edu.pl}
\author{Tomasz Gubiec\textsuperscript{1,4}}
\author{H. Eugene Stanley\textsuperscript{4}}
\affiliation{\textsuperscript{1}Institute of Experimental Physics,
  Faculty of Physics, University of Warsaw, Pasteura 5, 02-093 Warsaw,
  Poland} 
\affiliation{\textsuperscript{2}Instituto de Fisica Interdisciplinary
  Sistemas Complejos IFISC (CSIC-UIB), 07122 Palma de Mallorca, Spain} 
\affiliation{\textsuperscript{3}Scuola Normale Superiore, Piazza dei
  Cavalieri 7, 56126 Pisa, Italy} 
\affiliation{\textsuperscript{4}Center for Polymer Studies and
  Department of Physics, Boston University, Boston, Massachusetts 02215,
  USA} 

\maketitle

\section{Analytical approximation}

In the following we describe in detail our approach to approximating the
behavior of the coevolving spin model. The general form of the
Hamiltonian is
\begin{equation}\label{hamiltonian}
H(\{c_{ij}\},\{s_i\}) = - \sum_{i<j} c_{ij} \left( \frac{k_i
  k_j}{\langle  k \rangle} \right)^{\phi} s_i s_j - \sum_{i}
k_i^{\gamma}  - h \sum_{i} s_i, 
\end{equation}
where $s_i$ is the spin of node $i$ and $\langle k \rangle =
\frac{2M}{N} = 2c$ is the average degree of a network with $N$ nodes and
$M$ edges, described by the adjacency matrix $c_{ij}$. We set $h=0$, and
the model has two parameters $\gamma$ and $\phi$. To calculate the
energy and the maximum degree, which we use as a topological order
parameter, we calculate the partition function
\begin{equation}
Z = \sum_{\{c_{ij}\},\{s_i\}} e^{-\beta H(\{c_{ij}\},\{s_i\})}
\end{equation}
by differentiating the partition function
\begin{equation}
\frac{\partial Z}{\partial \beta} = - \sum_{\{c_{ij}\},\{s_i\}}
H(\{c_{ij}\},\{s_i\}) e^{-\beta H(\{c_{ij}\},\{s_i\})}.
\end{equation}
We obtain a formula for the energy of the system
\begin{equation}\label{energy}
- \frac{1}{Z} \frac{\partial Z}{\partial \beta} = -  \frac{\partial \ln
  Z}{\partial \beta} = E. 
\end{equation}

In the main text we focus on two cases, (i) when $\phi = 0$ and $\gamma$
can have any positive value, and (ii) when $\gamma = 1$, which is
equivalent to $\gamma = 0$, and a positive $\phi$ varies. In both cases we derive analytical approximations, which we describe below.

Note that when $\phi=0$ and $\gamma=0$ (or 1) we obtain a standard
Ising-like Hamiltonian with a coevolving network (see
Ref.~\cite{biely2009socio}). Note also that when $\phi=0$ and $\gamma=2$
we obtain a form of the Hamiltonian similar to that analyzed in
Refs.~\cite{palla2004statistical,berg2002correlated}, although those
references do not consider node states.

\subsection{First case: $\phi = 0$}\label{gamma}

\begin{figure}
\subfigure[~Model: $\phi = 0$, star nodes are in blue]{
 \includegraphics[width=0.35\linewidth]{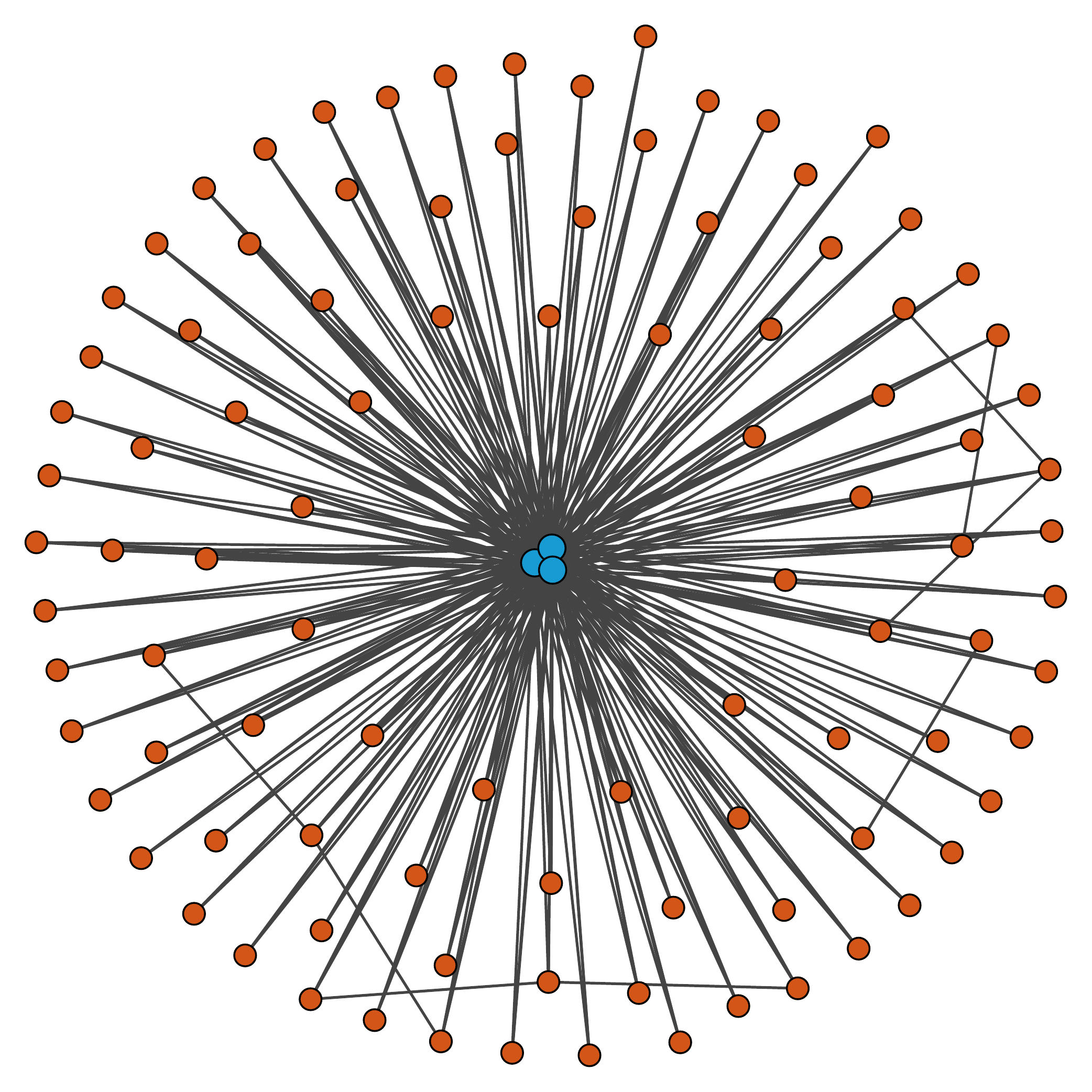}}
\hspace{1cm}
\subfigure[~Model: $\gamma = 1$, active component is in blue]{
 \includegraphics[width=0.35\linewidth]{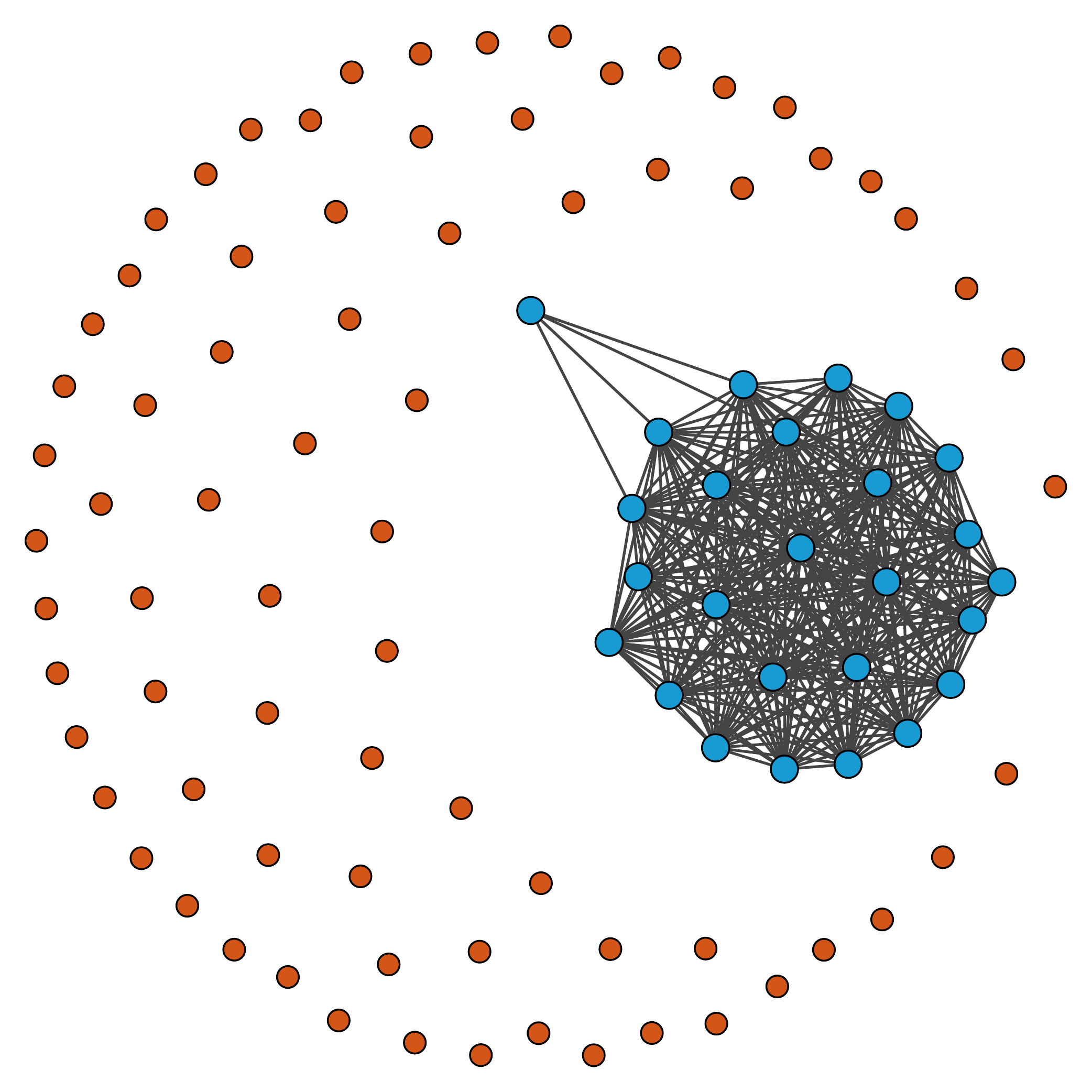}}
\caption{(Color online) Visualization of the configurations assumed for low
  temperatures in analytical approximations.}
\label{fig:graph}
\end{figure}

When $\phi = 0$ we simplify the Hamiltonian to be
\begin{equation}
H(\{c_{ij}\},\{s_i\}) = - \sum_{i<j} c_{ij} s_i s_j - \sum_{i} k_i^{\gamma} .
\end{equation}
Because the network can be strongly heterogeneous and the contribution
of highly connected nodes non-negligible, we cannot use a simple
mean-field approach, and we assume that the Hamiltonian is dominated by
the largest hubs with connections to all other nodes [see
  Fig.~\ref{fig:graph}(a)]. We also assume that all the other vertices
have the same degree $k_a$, and we neglect any effect of spins. This
gives us a simplified Hamiltonian
\begin{equation}
H \approx - n_h k_m^{\gamma} - (N-n_h)k^{\gamma}_a,
\end{equation}
where $n_h$ is the number of hubs with the maximum possible degree $k_m
= N-1$. Note that $n_h \leq \lfloor c \rfloor = \lfloor M/N \rfloor$. We
explain the derivation of the average degree $k_a$ for non-star vertices
below.  The hubs absorb a number of connections equal to
\begin{equation} 
L(n_h) = k_m + k_m-1 + ... + k_m - (n_h-1) = n_h
\frac{k_m+k_m-(n_h-1)}{2} = n_h \frac{2k_m+1-n_h}{2} = n_h
\frac{2N-1-n_h}{2}. 
\end{equation}
The remaining connections are distributed among non-star nodes, giving
them an average degree $k_a \equiv k_a(n_h) = n_h + \frac{2(M -
  L(n_h))}{N - n_h}$. We then approximate the number of connection
configurations among non-star vertices as a function of the number of
hubs using
\begin{equation}
R(n_h) = \binom{(N-n_h)(N-n_h-1)/2}{M-L(n_h)}.
\end{equation}
With the exception of when $n_h = \lfloor c \rfloor$, this approximation
overestimates the number of configurations for all other $n_h$ values,
but these overestimations are negligible.  Merging all of these
elements, the partition function is now
\begin{equation}
Z = 2^N  \sum_{n_h=0}^{c} \binom{N}{n_h} R(n_h) e^{\beta[ n_h k_m^\gamma
    + (N-n_h)k^{\gamma}_a]}, 
\end{equation}
where $2^N$ is the number of spin configurations.  Using
Eq.~(\ref{energy}) we calculate the energy,
\begin{equation}
\begin{split}
E = - \frac{1}{Z} \frac{\partial Z}{\partial \beta} = - \frac{2^N}{Z}
\sum_{n_h=0}^{\lfloor c \rfloor} \binom{N}{n_h} R(n_h) \left[ n_h
  k_m^\gamma + (N-n_h)k^{\gamma}_a\right] e^{\beta [ n_h k_m^\gamma +
    (N-n_h)k^{\gamma}_a]} = 
\\ = - \frac{2^N}{Z} \sum_{n_h=0}^{\lfloor c \rfloor} \binom{N}{n_h}
R(n_h) \left[ n_h (N-1)^\gamma + (N-n_h)k^{\gamma}_a\right] e^{\beta [
    n_h (N-1)^\gamma + (N-n_h)k^{\gamma}_a]} =  
\\ = - \frac{2^N}{Z} \Biggl[ (N-1)^\gamma \sum_{n_h=0}^{\lfloor c
    \rfloor} \binom{N}{n_h} R(n_h) n_h e^{\beta [ n_h (N-1)^\gamma +
      (N-n_h)k^{\gamma}_a]} + 
\\ + \sum_{n_h=0}^{\lfloor c \rfloor} \binom{N}{n_h} R(n_h)
(N-n_h)k^{\gamma}_a e^{\beta [ n_h (N-1)^\gamma + (N-n_h)k^{\gamma}_a]}
\Biggr]. 
\end{split}
\end{equation}
To determine the formula for the average maximum degree we first
describe its behavior for different $n_h$ values. When $n_h > 0$ there
is a minimum of one star, and thus the maximum network degree is $N-1$.
When $n_h=0$ there is no star, and we assume the network to be random.
Because a random network has an approximate Poisson degree distribution,
the maximum degree $\bar{k}(n_h)$ is
\begin{equation}
\bar{k}(n_h) = \left\{ \begin{array}{ll}
N-1 & \textrm{for $n_h>0$}\\
F_{Poiss,\lambda}^{-1}(1-\frac{1}{N}) & \textrm{for $n_h=0$}
\end{array} \right. ,
\end{equation}
where $F_{Poiss,\lambda}^{-1}$ is the inverse of a Poisson cumulative
distribution function with a parameter $\lambda = \langle k \rangle =
2M/N$. This yields the formula for the average maximum degree
\begin{equation}
\begin{split}
\langle k_{\max} \rangle &= \frac{2^N}{Z} \sum_{n_h=0}^{\lfloor c
  \rfloor} R(n_h) \bar{k}(n_h) e^{\beta [ n_h k_m^\gamma +
    (N-n_h)k^{\gamma}_a]} = \\ 
&= \frac{2^N}{Z} \left\{ R(0) \cdot
F_{Poiss,\lambda}^{-1}\left(1-\frac{1}{N}\right)  e^{\beta N
  (2M/N)^{\gamma}} + (N-1) \sum_{n_h=1}^{\lfloor c \rfloor} R(n_h)
e^{\beta [ n_h (N-1)^\gamma + (N-n_h)k^{\gamma}_a]} \right\}, 
\end{split}
\end{equation}
where $k_a$ is also a function of $n_h$, as defined above.

\subsection{Second case: $\gamma = 1$}\label{phi}

In the second case the Hamiltonian is equal to
\begin{equation}
H(\{c_{ij}\},\{s_i\}) = - \sum_{i<j} c_{ij} \left( \frac{k_i
  k_j}{\langle k \rangle} \right)^\phi s_i s_j. 
\end{equation}
We here approximate two-phase diagram effects, (i) the transition
between the ordered phase of a shattered network with one strongly
connected subgraph that contains all the edges and the disordered phase
of a random network, and (ii) the transition that occurs when $\phi =
\phi_c$ in the low-temperature regime. Here the network transitions from
an ordered structure to a structure with a few hubs (stars) that absorb
most of the connections, similar to the ordered phase described in
subsection \ref{gamma}.

\subsubsection{Active component approximation}

Figure~\ref{fig:graph}(b) shows an approximation that allows us to
describe analytically the transition from a shattered network with one
active, highly connected component to a random graph.  This approach
observes systemic behavior at different temperatures. Increasing the
temperature above zero causes the active component to attract
disconnected inactive nodes. Further increasing the temperature increases
the size of the active component, which can continue to grow until it encompasses
all nodes and becomes a random graph. Because we assume that every active component size can be approximated by a random graph with a given average degree, we use a mean-field description limited to the active nodes.  This gives us the Hamiltonian
\begin{equation}
H \approx - M \frac{\langle k \rangle_s^{2\phi}}{ \langle k
  \rangle^\phi}, 
\end{equation}
where $\langle k \rangle_s$ is the average degree of the active
component. When $n_s$ is the number of component nodes, $n_s \in \{
\lceil n_{\min} \rceil, \lceil n_{\min} \rceil + 1, \dots, N \}$, where
$n_{\min} = (1+\sqrt{1+8M})/2$, because all connections must fit in the
active component, and the limit is a complete subgraph with
$n_s(n_s-1)/2$ links. Because $\langle k \rangle_s = 2M/n_s$, we get
\begin{equation}
H \approx - 2^{\phi} N^{\phi} M^{\phi + 1} n_s^{-2\phi}.
\end{equation}
Note that as $n_s$ grows, the term ``component'' becomes misleading.  In
graph theory ``component'' is used to describe a subset of nodes
connected to each other. When $n_s$ becomes sufficiently large the
active component can break into two or more separate components.  We use
the term {\it active component\/} to mean a set of {\it active\/} nodes
able to obtain connections, in contrast to {\it inactive\/} nodes.

The number of configurations of connections for a given size of the
active component is equal to
\begin{equation}
R(n_s) = \binom{\frac{n_s(n_s-1)}{2}}{M}.
\end{equation}
The partition function is then
\begin{equation}
Z = \sum_{n_s = \lceil n_{\min} \rceil}^N 2^{N-n_s+1} \binom{N}{n_s}
R(n_s) e^{\beta \, 2^\phi \, N^\phi \, M^{\phi + 1} \, n_s^{-2\phi}}. 
\end{equation}
Finally we calculate the energy using Eq.~(\ref{energy}),
\begin{equation}
E = - \frac{1}{Z} \sum_{n_s = \lceil n_{\min} \rceil}^N 2^{\phi + N -
  n_s + 1} \, N^\phi \, M^{\phi + 1} \, n_s^{-2\phi} \binom{N}{n_s}
\binom{\frac{n_s(n_s-1)}{2}}{M} e^{\beta \, 2^\phi \, N^\phi \, M^{\phi
    + 1} \, n_s^{-2\phi}}. 
\end{equation}
Because we assume the active component is a random graph, we can
approximate its maximum degree in the same way as in subsection
\ref{gamma} for a zero-star configuration, but note that the maximum
degree of the components is limited by $n_s - 1$.  This is particularly
important when $n_s$ is small and the active component dense.  To avoid
overestimating the maximum degree, we write its dependence on $n_s$
\begin{equation}
\bar{k}(n_s) = \min\left\{n_s - 1,
F_{Poiss,\lambda_s}^{-1}\left(1-\frac{1}{n_s}\right)\right\}, 
\end{equation}
where $\lambda_s = \langle k \rangle_s = 2M/n_s$.  We use this formula
to obtain the average maximum degree
\begin{equation}
\begin{split}
\langle k_{\max} \rangle = \frac{2^N}{Z} \sum_{n_s = \lceil n_{\min}
  \rceil}^N \bar{k}(n_s) \, 2^{N - n_s + 1}   \binom{N}{n_s}
\binom{\frac{n_s(n_s-1)}{2}}{M} e^{\beta \, 2^\phi \, N^\phi \, M^{\phi
    + 1} \, n_s^{-2\phi}}. 
\end{split}
\end{equation}
This approximation also neglects the impact of spins.  This is
reasonable inside the two marginal phases, but inappropriate between
them.

\subsubsection{Critical value $\phi_c$}

We now estimate the critical line $\phi_c$ that at low temperatures
separates the model into two phases.  The phase for $\phi < \phi_c$ has
a fully connected cluster of size $n$ and many single disconnected
nodes. The real size of $n$ is the lower bound of $n_s$ introduced
above, and we approximate it using a continuous number $n \approx
n_{min} \equiv (1+\sqrt{1+8M})/2$. We here assume that all nodes in the
cluster are connected and all have the same degree $n-1$.  The energy of
the system is then
\begin{equation}\label{e1}
\begin{split}
E_{\phi < \phi_c} = - \sum^{\frac{n(n-1)}{2}}_{i=1} \left( \frac{2M}{N}
\right)^{-\phi} (n-1)^{2\phi}  =  
- \frac{n(n-1)}{2} \left( \frac{2M}{N} \right)^{-\phi} (n-1)^{2\phi} = -
M \left( \frac{2M}{N} \right)^{-\phi} (n-1)^{2\phi}. 
\end{split}
\end{equation}
Nodes separated from the active component are zero degree and do not
contribute to the energy.

When $\phi > \phi_c$ there are dominant hubs of maximum degree $N-1$.
We approximate the energy of the system by taking into account
interactions between $n_h$ stars and between stars and all other $N-n_h$
nodes. We omit interactions between non-star nodes, where the degree
multiplication is significantly smaller. We also assume non-star nodes
have the same average degree.  Finally we get
\begin{equation}\label{e2}
\begin{split}
E_{\phi > \phi_c} = - \sum^{\frac{n_h(n_h-1)}{2}}_{i=1} \left(
\frac{2M}{N} \right)^{-\phi} (N-1)^{2\phi} - \sum^{n_h(N-n_h)}_{i=1}
\left( \frac{2M}{N} \right)^{-\phi} (N-1)^{\phi} n_h^{\phi} = 
\\ = - \frac{n_h(n_h-1)}{2} \left( \frac{2M}{N} \right)^{-\phi}
(N-1)^{2\phi} - n_h (N-n_h) \left( \frac{2M}{N} \right)^{-\phi}
(N-1)^{\phi} n_h^{\phi} =  
\\ = - n_h (N-1)^{\phi} \left( \frac{2M}{N} \right)^{-\phi} \left[
  \frac{n_h-1}{2} (N-1)^{\phi} + n_h^{\phi} 
(N-n_h) \right],
\end{split}
\end{equation}
where $n_h$ is the number of hubs.  When $T \rightarrow 0$ we have $n_h
\approx 2M/N = c$ we assume it is continuous.  At the critical line both
energies (\ref{e1}) and (\ref{e2}) are equal
\begin{equation}
\begin{split}
c (N-1)^{\phi_c} \left( \frac{2M}{N} \right)^{-\phi_c} \left[
  \frac{c-1}{2} (N-1)^{\phi_c} + c^{\phi_c} (N-c) \right] =  M \left(
\frac{2M}{N} \right)^{-\phi_c} (n-1)^{2\phi_c} , \\ 
\frac{c-1}{2} (N-1)^{\phi_c} + c^{\phi_c} (N-c) = \frac{M}{c}
\left(\frac{(n-1)^2}{N-1}\right)^{\phi_c}, \\ 
 \ln \left( \frac{c-1}{2} (N-1)^{\phi_c} + c^{\phi_c} (N-c) \right) =
 \phi_c \ln \frac{(n-1)^2}{N-1} + \ln \frac{M}{c} . 
\end{split}
\end{equation}
This allows us to numerically approximate $\phi_c$.

\section{Simulation}

We use the Metropolis algorithm
\cite{metropolis1953equation,beichl2000metropolis,wolfram2002new} to perform Monte Carlo simulations of the model. We begin every simulation with a random graph and a random spin configuration.  Before collecting data about the system, we run the simulation for a given thermalization time determined by network size and the parameter values. We calculate the thermalization by observing the time evolution of order parameters and find it equal to the time needed to reach a stationary state with values fluctuating around the equilibrium. We then collect the order parameter values for a number of time steps.

Every time step of the simulation uses two basic mechanisms, (i) spin switching and (ii) edge rewiring.  In spin switching, we randomly select one node and compute the energy difference between the current state of the system and a chosen spin in the opposite state. Note that here the energy difference is a function of the selected spin and its neighbors. Because here the network structure is constant, we take into account only the first Hamiltonian sum (\ref{hamiltonian}). We use the standard Metropolis rule to decide whether to flip the spin.

When we perform edge rewiring at the same time step, we randomly select one network link and two nodes to serve as possible ends of a new edge. Note that the new edge cannot overlap with any existing edges. We next calculate the energy difference between the new configuration (with the new link and without the old link) and the old configuration.  Here the energy difference depends on all four (or in some rare cases three) nodes connected by the old and new connections. The difference is determined by the neighboring nodes.  Also, both the pure topological and mixed terms of the Hamiltonian are influenced. We again use the Metropolis algorithm to decide whether to rewire the edge.

We use a $1:1$ ratio between the spin switching and the link rewiring to minimize thermalization times.  We also perform simulations using other ratios. We find that only difference is in the time necessary to reach the stationary state. When ratios are extreme, the thermalization times become infinite. In pure spin switching we find a constant network structure, and in pure link rewiring the magnetization does not change. As long as the ratio is finite, the simulation trajectories cover the entire configuration space, and the equilibrium states remain the same.

To better understand and to easily reproduce and extend our results, we present a pseudo-code of the main loop of our simulation.  It supplies the procedures performed in every time step.

\begin{verbatim}
graph = random_Erdos_Renyi_graph(n, m)  // generate random graph
for i in [0, ..., n-1]:
    graph[i][spin] = random_chice(-1, +1)  // assign random spin to every node
    
for step in time_steps:
    v_index = rand_int(n)  // chose one of n nodes
    e_index = rand_int(m)  // chose one of m edges

    // spin switching
    neighbors = graph.get_neighbors(v_index)
    delta = graph.energy_change_spin(v_index, neighbors)  // compute energy difference
    if delta <= 0 or rand() < exp(- delta / T):
        graph[i][spin] = graph[i][spin] * -1  // flip spin according to Metropolis rule

    // edge rewiring
    old_from, old_to = graph.get_edge_ends(e_index)
    new_from, new_to = graph.draw_new_edge(exclude_existing=true)
    // compute energy difference
    delta = graph.energy_change_edge(old_from, old_to, new_from, new_to)
    if delta <= 0 or rand() < exp(- delta / T):
        graph.delete_edge([v_from, v_to])  // rewire edge according to Metropolis rule
        graph.add_edge([new_from, new_to])
\end{verbatim}